\documentclass[aps,pra,twocolumn,showpacs]{revtex4}
\usepackage{amsmath}

\begin{document}
\title{Quantum theory of successive projective measurements}

\author{Lars M. Johansen}
\affiliation{Department of Technology, Buskerud University College,
N-3601 Kongsberg, Norway}
\date{\today}
\email{lars.m.johansen@hibu.no}
\begin{abstract}

We show that a quantum state may be represented as the sum of a
joint probability and a complex quantum modification term. The joint
probability and the modification term can both be observed in
successive projective measurements. The complex modification term is
a measure of measurement disturbance. A selective phase rotation is
needed to obtain the imaginary part. This leads to a complex
quasiprobability, the Kirkwood distribution. We show that the
Kirkwood distribution contains full information about the state if
the two observables are maximal and complementary. The Kirkwood
distribution gives a new picture of state reduction. In a
nonselective measurement, the modification term vanishes. A
selective measurement leads to a quantum state as a nonnegative
conditional probability. We demonstrate the special significance of
the Schwinger basis.

\end{abstract}


\maketitle

\section{Introduction}

Quantum mechanics is a probabilistic theory, in the sense that all
experimental predictions are probabilistic. However, it uses
concepts unfamiliar from the classical theory of probability. The
concept of a quantum state has only partially a direct
interpretation in terms of probabilities. A pure state is a unit
vector in a complex Hilbert space, and a mixed state is a positive
definite hermitian matrix of trace one. Remarkably, the probability
interpretation of quantum mechanics was introduced as a footnote
added in proof \cite{Born-QuanStos:26}.

On the other hand, a state in classical physics may be represented
as a joint probability over a classical logic of proposals. The
proposals are basic elements of phase space. This state concept is
too narrow to encompass quantum mechanics, as witnessed by the
theorems of Bell \cite{Bell-EinsPodoRosepara:64} and Kochen and
Specker \cite{Kochen+Specker-ProbHiddVariQuan:67}. The element in
quantum mechanics that corresponds to the classical proposal is the
projector. The essential difference between quantum and classical
physics is that quantum proposals do not commute.

Is there any connection between the state concepts in classical and
quantum physics? We demonstrate in this paper that a quantum state
may be represented as the sum of a nonnegative joint probability and
a quantum modification term. Thus, we represent quantum states as a
modification of the classical state concept, i.e. as a
\emph{quasiprobability}. Nonclassical values of this
quasiprobability are due to measurement disturbance. The complex
quasiprobability that we obtain was first discovered by Kirkwood as
a representation of quantum states over phase space
\cite{Kirkwood-QuanStatAlmoClas:33}. We shall refer to it as the
Kirkwood distribution. It was independently rediscovered and
generalized to arbitrary observables by Dirac
\cite{Dirac-AnalBetwClasQuan:45} and Barut
\cite{Barut-DistFuncNoncOper:57}. The real part of this distribution
was examined in phase space by Terletsky
\cite{Terletsky-claslimiquanmech:37} and for arbitrary pairs of
observables by Barut \cite{Barut-DistFuncNoncOper:57} and Margenau
and Hill \cite{Margenau+Hill-CorrbetwMeasQuan:61}. The question of a
possible connection between the Kirkwood distribution and
measurement disturbance was raised by Prugove\v{c}ki
\cite{Prugovecki-TheoMeasIncoObse:67}.

The concept of a phase space has been imported from classical
physics into quantum mechanics. There are many representations of
quantum states in terms of phase space distributions
\cite{Wigner-QuanCorrTherEqui:32,%
Kirkwood-QuanStatAlmoClas:33,Husimi-SomeFormPropDens:40,%
Sudarshan-EquiSemiQuanMech:63,Cohen-GenePhasDistFunc:66,%
Agarwal+Wolf-CalcFuncNoncOper:70}. Most popular among these is the
Wigner distribution \cite{Wigner-ProbMeas:63}. This is a real
distribution, but it may take negative values.

It is not possible to establish a joint probability in the classical
sense for noncommuting observables
\cite{Neumann-MathFounQuanMech:55}. However, one may establish
quasiprobabilities for such observables
\cite{Dirac-AnalBetwClasQuan:45}. A fundamental question that
remains unanswered is which observables, except for the standard
phase space observables, that may provide a complete state
description in terms of a quasiprobability.

There exists various generalizations of the Wigner distribution
concept beyond the standard continuous phase space (see e.g. Refs.
\cite{Moyal-QuanMechStatTheo:49,Stratonovich-DistReprSpac:57,%
Wootters-WignFormFiniQuan:87,Galetti+Pisa-ExteWeylTranSpec:88,%
Leonhardt-QuanTomoDiscWign:95}). These distributions are mostly
different, since there exists no general agreement on how the Wigner
distribution concept is to be generalized. For systems in a finite
dimensional Hilbert space, the issue of informational completeness
has been dealt with in particular by Wootters
\cite{Wootters-WignFormFiniQuan:87} and Leonhardt
\cite{Leonhardt-QuanTomoDiscWign:95}. In infinite dimensional
Hilbert space, informationally complete generalizations of the
Wigner distribution have been found for photon number and phase
\cite{Luks+Perinova-OrdeLaddOperWign:93,Vaccaro-NumbWignfuncFock:95}.

We demonstrate in this paper that the Kirkwood distribution over a
pair of observables determines the density matrix uniquely provided
that the observables are maximal and complementary. A maximal
observable has a nondegenerate spectrum. We define two observables
as complementary if they have no common eigenvectors, i.e. there is
no state where both observables have a well-defined value. This
generalizes the concept of informationally complete quasiprobability
distributions considerably.

An issue that has been at the center of discussions over the
foundations of quantum mechanics for very long time is the question
of state reduction or collapse. We shed some new light on this issue
in this paper by rephrasing the standard formalism of state
reduction \cite{Neumann-MathFounQuanMech:55,Lueders-UEbeZust:51} in
terms of the Kirkwood distribution. We show that nonselective state
reduction, which is represented in the standard representation as a
vanishing of offdiagonal elements of the density matrix, is
represented here as a vanishing of the complex modification term.
Furthermore, selective state reduction, which is sometimes called
the ``collapse'', is here represented merely as a conditional
probability. The latter result is in correspondence with results
based on other approaches to this problem
\cite{Bub-NeumProjPostProb:77}.

Projective measurements were the only type of measurements
considered in orthodox quantum theory
\cite{Dirac-PrinQuanMech:58,Neumann-MathFounQuanMech:55}. Projective
measurements require the preparation of the measurement apparatus in
a state with a sharp position of the pointer prior to the
measurement interaction \cite{Neumann-MathFounQuanMech:55}. Aharonov
\emph{et al.} \cite{Aharonov+AlbertETAL-ResuMeasCompSpin:88} found a
certain regularity in measurements where the measurement apparatus
would be prepared in a state with an uncertain pointer position. In
particular, they found that successive measurements of incompatible
observables using this form of measurement interaction for the first
observable gave results that they called ``weak values''. This type
of measurement was soon found to give the first operational
significance to the Kirkwood distribution, since the weak values
could be expressed as conditional expectation values of the Kirkwood
distribution \cite{Steinberg-Condprobquantheo:95}. It was recently
shown that weak values also may be reconstructed from projective
measurements \cite{Johansen-Recoweakvaluwith:07}. In this paper, we
explore the operational significance of the Kirkwood distribution in
terms of projective measurements.

This paper is organized as follows. In Sec. \ref{sec:successive} we
introduce the necessary quantum formalism. We review the standard
formalism of state reduction in successive measurements and the
derivation of the Wigner formula \cite{Wigner-ProbMeas:63}. In Sec.
\ref{sec:quantum} we show that the Kirkwood distribution is obtained
as a modification of the Wigner formula. We also represent the
standard formalism of state reduction in terms of the Kirkwood
distribution. In Sec. \ref{sec:completeness} we determine necessary
and sufficient conditions for the informational completeness of the
Kirkwood distribution. We also demonstrate the particular
significance taken by the Schwinger basis
\cite{Schwinger-UnitOperBase:60}.

\section{Successive measurements and the Wigner formula}
\label{sec:successive}

In this section, we establish the notation and review known results
concerning successive projective measurements. We define projective
measurements in the standard way as defined for observables with
nondegenerate spectra by von Neumann
\cite{Neumann-MathFounQuanMech:55} and as generalized to observables
with degenerate spectra by L\"{u}ders \cite{Lueders-UEbeZust:51}.

We consider two observables $\hat{A}$ and $\hat{B}$ with spectral
resolutions
\begin{subequations}
\begin{eqnarray}
    \label{eq:obsA}
    \hat{A} &=& \sum_m a_m \hat{A}_m, \\
    \label{eq:obsB}
    \hat{B} &=& \sum_n b_n \hat{B}_n,
\end{eqnarray}
\end{subequations}
where $a_m$ $(m=1,2,...N_a)$ and $b_n$ $(n=1,2,...N_b)$ are
eigenvalues and $\hat{A}_m$ and $\hat{B}_n$ are the corresponding
eigenprojectors. $N_a\le N$ and $N_b\le N$, where $N$ is the
dimension of the Hilbert space. These eigenvalues form a rectangular
lattice of dimensions $N_a \times N_b$. For the case of
nondegenerate eigenvalues, $N_a=N_b=N$.

The projectors are assumed to be orthogonal and idempotent,
\begin{subequations}
\begin{eqnarray}
    \label{eq:orthoidemA}
    \hat{A}_m \hat{A}_n &=& \delta_{mn} \hat{A}_n, \\
    \hat{B}_m \hat{B}_n &=& \delta_{mn} \hat{B}_n.
\end{eqnarray}
\end{subequations}
Eigenvalues may be degenerate, i.e., we have
\begin{subequations}
\begin{eqnarray}
    \mathrm{Tr} \hat{A}_m &\ge& 1,\\
    \mathrm{Tr} \hat{B}_n &\ge& 1.
\end{eqnarray}
\end{subequations}
The eigenprojectors sum to unity, i.e., we have the projection
valued measures
\begin{subequations}
\begin{eqnarray}
    \label{eq:pvmA}
    \sum_m \hat{A}_m &=& \hat{1}, \\
    \label{eq:pvmB}
    \sum_n \hat{B}_n &=& \hat{1}.
\end{eqnarray}
    \label{eq:PVM}
\end{subequations}
We now consider projective measurements on a system prepared in the
state $\hat{\rho}$. According to Born's postulate, the probability
of obtaining the eigenvalue $a_m$ in a projective measurement of the
observable $\hat{A}$ is
\begin{equation}
    P(a_m) = \mathrm{Tr} \hat{\rho} \hat{A}_m.
\end{equation}
This is also the probability of obtaining the value 1 in a
projective measurement of the projector $\hat{A}_m$. Note that these
are two different ways of measuring this probability.

Since we want to study successive measurements, we also need to take
into consideration how the system is affected by the first
measurement. This may be done by including into the description a
model of the measuring device. Here we will take a simplified
approach and represent the state change by the projection postulate
\cite{Neumann-MathFounQuanMech:55}. The equivalence of these two
methods have been demonstrated for certain models of the measurement
apparatus (see, e.g., Refs.
\cite{Daneri+LoingerETAL-QuanTheoMeasErgo:62,
Hepp-QuanTheoMeasMacr:72}). We will use the projection postulate in
the form proposed by L\"{u}ders \cite{Lueders-UEbeZust:51} for
observables with possibly degenerate spectra. This type of state
reduction has been demonstrated to follow from the requirement of
the measurement being repeatable and minimally disturbing
\cite{Goldberger+Watson-MeasTimeCorrQuan:64,%
Bell+Nauenberg-MoraAspeQuanMech:66,Herbut-DeriChanStatMeas:69}.

We divide projective measurements into two types, selective and
nonselective. They can be made with the same measurement
interaction. The distinction only applies to the way we treat the
ensemble \emph{after} the measurement interaction is over. A
selective measurement is one where we only keep the subensemble
giving a particular outcome. In a nonselective measurement, the
complete ensemble that was prepared initially is considered further
as a whole.

The state after the measurement will depend on whether we perform a
measurement of the complete observable $\hat{A}$ or whether we
perform a measurement of just one projector $\hat{A}_m$. We shall
consider here the case of a projector measurement. This is an
experiment with only two possible outcomes. For a nonselective
projective measurement of the L\"{u}ders type of an projector
$\hat{A}_m$ of the form (\ref{eq:obsA}), the post-measurement state
is \cite{Lueders-UEbeZust:51},
\begin{equation}
    \hat{\rho}' = \hat{A}_m \hat{\rho} \hat{A}_m + (\hat{1} -
    \hat{A}_m) \hat{\rho} (\hat{1} - \hat{A}_m),
    \label{eq:nonselective}
\end{equation}
where $\hat{1}-\hat{A}_m$ is the orthogonal complement to
$\hat{A}_m$,
\begin{equation}
    \hat{A}_m (1-\hat{A}_m) = (1-\hat{A}_m) \hat{A}_m = 0.
\end{equation}
In case the projector measurement gave as a result the value 1, the
selective post-measurement state is \cite{Lueders-UEbeZust:51}
\begin{equation}
    \hat{\rho}_s' = {\hat{A}_m \hat{\rho} \hat{A}_m \over
    \mathrm{Tr} \hat{\rho} \hat{A}_m}.
    \label{eq:selective}
\end{equation}
For a rank one projector $\hat{A}_m=| a_m \rangle\langle a_m |$, the
selective post-measurement state is $\hat{\rho}_s'= \hat{A}_m$. This
is sometimes referred to as ``collapse''. In this case, the initial
state is completely erased.

We now assume that a projective measurement of the observable
$\hat{B}$ is made on the system after a selective measurement of
$\hat{A}_m$. The probability of obtaining the eigenvalue $b_n$ on
the selective state $\hat{\rho}_s'$ is then \cite{Johansen-Foot}
\begin{equation}
    P(b_n | a_m) = \mathrm{Tr} \hat{\rho}_s' \hat{B}_n = {\mathrm{Tr}
    \hat{\rho} \hat{A}_m \hat{B}_n \hat{A}_m \over \mathrm{Tr}
    \hat{\rho} \hat{A}_m}.
\end{equation}
The joint probability of obtaining successively the eigenvalues
$a_m$ and $b_n$ is therefore
\begin{equation}
    P(a_m) P(b_n | a_m) = \mathrm{Tr} \hat{\rho} \hat{A}_m
    \hat{B}_n \hat{A}_m.
    \label{eq:wignerformula}
\end{equation}
This probability for successive measurements was introduced for
observables with nondegenerate spectra by Wigner
\cite{Wigner-ProbMeas:63}. For this reason, it is sometimes referred
to as the Wigner formula. Of course, this is not a joint probability
in the classical sense, since reversing the order of operations does
not lead to the same probability.

The marginal probabilities obtained from the this joint probability
are
\begin{subequations}
\begin{eqnarray}
    \sum_n \mathrm{Tr} \hat{\rho} \hat{A}_m \hat{B}_n \hat{A}_m &=&
    \mathrm{Tr} \hat{\rho}  \hat{A}_m, \\
    \sum_m \mathrm{Tr} \hat{\rho} \hat{A}_m \hat{B}_n \hat{A}_m &=&
    \mathrm{Tr} \hat{\rho}' \hat{B}_n.
    \label{eq:WBmarg}
\end{eqnarray}
    \label{eq:wignermarginals}
\end{subequations}
We see that the marginal probability for $\hat{A}_m$ is related to
the pre-measurement state $\hat{\rho}$. However, the marginal
probability for $\hat{B}_n$ is given in terms of the nonselective
post-measurement state $\hat{\rho}'$. Thus, the two marginals relate
to different states. In this sense, Wigner's formula expresses
properties of both the pre- and the post-measurement state. One may
therefore understand that it does not contain complete information
about the pre-measurement state $\hat{\rho}$.

\section{The Kirkwood distribution}
\label{sec:quantum}

In this section, we investigate the operational significance of the
Kirkwood distribution. In the most general form, this distribution
is written simply as \cite{Kirkwood-QuanStatAlmoClas:33,%
Dirac-AnalBetwClasQuan:45,Barut-DistFuncNoncOper:57}
\begin{equation}
    P(a_m,b_n) = \mathrm{Tr} \hat{\rho} \hat{A}_m \hat{B}_n.
\end{equation}
Although it is complex, it follows trivially from the completeness
relations (\ref{eq:PVM}) that it gives correct marginal
probabilities,
\begin{subequations}
\begin{eqnarray}
    \sum_n \mathrm{Tr} \hat{\rho} \hat{A}_m \hat{B}_n &=&
    \mathrm{Tr} \hat{\rho} \hat{A}_m,\\
    \sum_m \mathrm{Tr} \hat{\rho} \hat{A}_m \hat{B}_n &=&
    \mathrm{Tr} \hat{\rho} \hat{B}_n.
\end{eqnarray}
\end{subequations}
We again consider a projective measurement of the projector
$\hat{A}_m$ as described in the previous section. We expand the
expression for the nonselective post-measurement state
(\ref{eq:nonselective}), multiply both sides with $\hat{B}_n$, and
compute the trace. After a rearrangement of terms, we arrive at the
expression
\begin{equation}
    \mathrm{Re} (\mathrm{Tr} \hat{\rho} \hat{A}_m \hat{B}_n) =
    \mathrm{Tr}\hat{\rho} \hat{A}_m \hat{B}_n \hat{A}_m +
    {1\over 2} \mathrm{Tr} \left (  \hat{\rho} - \hat{\rho}'
    \right ) \hat{B}_n.
    \label{eq:mhproj}
\end{equation}
The l.h.s. of this expression is the real part of the Kirkwood
distribution. It is often referred to as the Margenau-Hill
distribution \cite{Margenau+Hill-CorrbetwMeasQuan:61}. It may take
negative values, and it gives correct marginal probabilities
\cite{Margenau+Hill-CorrbetwMeasQuan:61}
\begin{subequations}
\begin{eqnarray}
    \sum_m \mathrm{Re} (\mathrm{Tr} \hat{\rho} \hat{A}_m \hat{B}_n)
    &=& \mathrm{Tr} \hat{\rho} \hat{B}_n,\\
    \sum_n \mathrm{Re} (\mathrm{Tr} \hat{\rho} \hat{A}_m \hat{B}_n)
    &=& \mathrm{Tr} \hat{\rho} \hat{A}_m.
\end{eqnarray}
\end{subequations}
We see from Eq. (\ref{eq:mhproj}) that the Margenau-Hill
distribution is expressed in terms of the Wigner formula
(\ref{eq:wignerformula}) and a second terms. The second term is
proportional to the difference in the expectation value of the
projector $\hat{B}_n$ measured on the initial state $\hat{\rho}$ and
on the nonselective state $\hat{\rho}'$ after a measurement of
$\hat{A}_m$. This is therefore proportional to the change imposed on
the probability of $\hat{B}_n$ due to an intervening measurement of
$\hat{A}_m$.

The second term on the r.h.s. of Eq. (\ref{eq:mhproj}), the quantum
modification term, is the cause of any negative values of the
Margenau-Hill distribution. It vanishes if the two projectors
$\hat{A}_m$ and $\hat{A}_n$ commute. Naturally, commuting projectors
are supported by a nonnegative joint probability. Also, it vanishes
for states that do not change during a measurement. For example, it
vanishes if (\ref{eq:nonselective}) is taken as the initial state.
We understand that the Margenau-Hill distribution may be
reconstructed from the various probabilities that are found on the
r.h.s. These probabilities are obtained from measurements of the two
projectors $\hat{A}_m$ and $\hat{B}_n$ only.

It may be shown that in general the Margenau-Hill distribution
(\ref{eq:mhproj}) does not determine the density matrix uniquely. It
will be shown in the next section that the Kirkwood-distribution
does determine the density matrix for a wide class of observables.
We shall therefore consider also the imaginary part of the
Kirkwood-distribution.

We start by introducing the operator
\begin{equation}
    \hat{R}_m^{\phi} = 1+(e^{i \phi} -1) \hat{A}_m,
\end{equation}
where $\phi$ is a real parameter. By Eq. (\ref{eq:orthoidemA}) we
have
\begin{equation}
    \hat{R}_m^{\phi} \hat{A}_n = 1+(e^{i \phi} -1) \delta_{mn}
    \hat{A}_n.
\end{equation}
$\hat{R}_m^{\phi}$ may be regarded as a selective phase rotation
operator. It affects only a single projector $\hat{A}_m$ in the
complete set (\ref{eq:pvmA}). $\hat{R}_m^{\phi}$ can be implemented
at time $t_0$ by adding to the Hamiltonian a term
\begin{equation}
    \Delta \hat{H}_m = - \phi \delta(t-t_0) \hat{A}_m.
\end{equation}
It may be noted that the nonselective post-measurement density
operator (\ref{eq:nonselective}) may be written as
\begin{equation}
    \hat{\rho}' = {1 \over 2} \left [ \hat{\rho} + \hat{R}_m^\pi
    \, \hat{\rho} \, (\hat{R}_m^\pi)^\dag \right ].
    \label{eq:random}
\end{equation}
This shows that the L\"{u}ders form of the projection postulate
\cite{Lueders-UEbeZust:51} leads to a phase randomization in the
measurement of any type of projector, regardless of degeneracy. It
may now be verified that
\begin{equation}
    \label{eq:im}
    \mathrm{Im} (\mathrm{Tr} \hat{\rho} \hat{A}_m \hat{B}_n) =
    {1 \over 2} \mathrm{Tr} \left ( \hat{\rho} - \hat{\rho}'
    \right ) \hat{B}_n^{\pi/2},
\end{equation}
where
\begin{equation}
    \hat{B}_n^{\pi/2} = \hat{R}_m^{\pi/2} \hat{B}_n
    (\hat{R}_m^{\pi/2})^\dag
\end{equation}
is a projector obtained by performing a selective phase rotation
$\hat{R}_m^{\pi/2}$ on the projector $\hat{B}_n$.

This shows that the imaginary part of the Kirkwood distribution is
obtained by observing the change in the expectation value of the
selectively phase rotated projector $\hat{B}_n^{\pi/2}$ due to an
intermediate projective measurement of the other projector
$\hat{A}_m$.

It follows from the completeness relation (\ref{eq:PVM}) that the
marginals of the imaginary part vanish,
\begin{subequations}
\begin{eqnarray}
    \sum_m \mathrm{Im} (\mathrm{Tr} \hat{\rho} \hat{A}_m \hat{B}_n)
    &=& 0,\\
    \sum_n \mathrm{Im} (\mathrm{Tr} \hat{\rho} \hat{A}_m \hat{B}_n)
    &=& 0.
\end{eqnarray}
\end{subequations}
Of course, the classical equivalent of the imaginary modification
term is vanishing.

It may be of interest to examine how the standard formalism of
projective measurements, and in particular the von
Neumann-L\"{u}ders rules of state reduction, are expressed in terms
of the Kirkwood distribution. A straightforward calculation shows
that the Kirkwood distribution for the nonselective post-measurement
state $\hat{\rho}'$ in (\ref{eq:nonselective}) is
\begin{equation}
    \label{eq:nonselectivedirac}
    \mathrm{Tr} \hat{\rho}' \hat{A}_m \hat{B}_n = \mathrm{Tr}
    \hat{\rho}  \hat{A}_m \hat{B}_n \hat{A}_m.
\end{equation}
We recognize this as the nonnegative Wigner formula
(\ref{eq:wignerformula}), i.e., the joint probability for successive
measurements. Thus, the effect of a nonselective projective
measurement is that the complex modification term vanishes. This is
usually associated with a vanishing of the off-diagonal elements of
the density matrix \cite{Neumann-MathFounQuanMech:55}.

Furthermore, we find that the Kirkwood distribution for the
selective post-measurement state (\ref{eq:selective}) is
\begin{equation}
    \label{eq:selectivedirac}
    \mathrm{Tr} \hat{\rho}_s' \hat{A}_m \hat{B}_n = {\mathrm{Tr}
    \hat{\rho}  \hat{A}_m \hat{B}_n \hat{A}_m \over
    \mathrm{Tr} \hat{\rho} \hat{A}_m}.
\end{equation}
This is just the conditional probability calculated from the
unconditional post-measurement probability
(\ref{eq:nonselectivedirac}). Thus, a selective measurement, often
represented as a ``collapse'' in the standard formulation, is just a
classical probability conditionalization in the Kirkwood
representation. This lends support to the notion of a collapse as a
statistical effect.

\section{Informational completeness}
\label{sec:completeness}

In this section, we demonstrate that the Kirkwood distribution
determines the density matrix uniquely for a wide class of
observables. Although this proof is rather trivial, we haven't been
able to find it explicitly in the literature, and it is included
here for completeness.

We introduce two complete orthonormal bases
\begin{subequations}
\begin{eqnarray}
    \sum_m | a_m \rangle \langle a_m | &=& \hat{1},\\
    \sum_n | b_n \rangle \langle b_n | &=& \hat{1}.
\end{eqnarray}
    \label{eq:nondegpvm}
\end{subequations}
Therefore, we restrict the attention to observables with
nondegenerate spectra. The density matrix in one basis may be
expressed in terms of the second basis as
\begin{equation}
    \langle b_m | \hat{\rho} | b_n \rangle = \sum_k \langle b_m |
    \hat{\rho} | a_k \rangle \langle a_k | b_n \rangle.
\end{equation}
By introducing the Kirkwood distribution
\begin{equation}
    \mathrm{Tr} \hat{\rho} | a_k \rangle \langle a_k
    | b_m \rangle \langle b_m | = \langle b_m |
    \hat{\rho} | a_k \rangle \langle a_k | b_m \rangle,
\end{equation}
we may write the density matrix in the form
\begin{equation}
    \langle b_m | \hat{\rho} | b_n \rangle = \sum_k {\langle a_k | b_n
    \rangle \over \langle a_k | b_m \rangle} \, \mathrm{Tr} \hat{\rho}
    | a_k \rangle \langle a_k | b_m \rangle \langle b_m |.
    \label{eq:inversion}
\end{equation}
This equation shows how the density matrix may be obtained by a
transformation of the Kirkwood distribution for a pair of
nondegenerate observables. The transformation (\ref{eq:inversion})
goes through for any bases if $\langle a_k | b_m \rangle \ne 0$ for
all $(m,n)$. If $\langle a_k | b_m \rangle = 0$ for at least one
pair of indices $(m,n)$, the corresponding term in the sum is
indeterminate. Thus, the Kirkwood distribution determines the
density matrix uniquely for any pair of orthonormal and mutually
nonorthogonal bases.

A complete orthonormal basis is maximal, in the sense that there are
no more vectors that are orthogonal to this basis. Thus, if a second
complete orthonormal basis has one vector which is normal to one
vector in the first basis, then this vector must also belong to the
first basis. This means that if the two bases have at least one pair
of mutually orthogonal vectors, then they must have at least one
vector in common. Thus, the requirement that the bases should have
no mutually orthogonal vectors is equivalent to the claim that they
should have no common vectors. Such bases are sometimes referred to
as complementary (see, e.g., Ref.
\cite{Beltrametti+Cassinelli-LogiQuanMech:81}). Such observables
cannot both have sharp values. Note that this requirement is
stronger than the requirement that the observables should be
noncommuting. Noncommuting observables may still have some
eigenvectors in common, and if a system is prepared in one of those
common eigenvectors, both noncommuting observables have sharp
values.

A more strict definition of complementary observables is that sharp
knowledge of one observable should imply that all values of the
other observable are equally probable. For an $N$-dimensional
Hilbert space, this implies that
\begin{equation}
    | \langle a_m | b_n \rangle | = {1 \over \sqrt{N}}
\end{equation}
for all $(m,n)$. Such bases have been called mutually unbiased
\cite{Wootters-QuanMechwithProb:86}. Schwinger proposed a particular
implementation of mutually unbiased bases as
\cite{Schwinger-UnitOperBase:60}
\begin{equation}
    \langle a_m | b_n \rangle =  {1 \over \sqrt{N}} \; e^{2 \pi i m n
    / N}.
    \label{eq:schwingerbasis}
\end{equation}
Discrete Wigner functions have been constructed using both the
Schwinger bases \cite{Galetti+Pisa-ExteWeylTranSpec:88} and more
general mutually unbiased bases \cite{Wootters-WignFormFiniQuan:87}.
For the Schwinger basis, the inversion formula (\ref{eq:inversion})
simplifies to
\begin{equation}
    \label{eq:fourier}
    \langle b_m | \hat{\rho} | b_n \rangle = \sum_k e^{2 \pi i k (n-m)/N
    } \mathrm{Tr} \hat{\rho} | a_k \rangle \langle a_k | b_m \rangle
    \langle b_m |.
\end{equation}
Thus, in this case the transformation between the Kirkwood
distribution and the density matrix is a discrete Fourier transform.

Since we are able to reconstruct the density matrix from
complementary bases, we may also transform the Kirkwood distribution
between different bases without first transforming to the density
matrix. This possibility has been explored in Ref.
\cite{Pimpale+Razavy-Quanphasspac:88}.

\section{Conclusion}

We have shown that the Kirkwood distribution generalizes the
classical concept of a state as a joint probability by adding a
complex modification term. The joint probability is the Wigner
formula, which is obtained in a successive measurement. The complex
modification term is obtained from a measurement of the measurement
disturbance. We needed the disturbance on the projector itself and
on a selectively phase rotated projector.

We demonstrated that the Kirkwood distribution gives a complete
description of a quantum state provided that the two observables
have nondegenerate spectra and no common eigenvectors. This
considerably enlarges the class of informationally complete
quasiprobabilities.

We demonstrated that state reduction due to projective measurements
gives a different perspective to quantum measurement theory. We
found that in a nonselective measurement the complex modification
term vanishes, and the quasiprobability reduces to the Wigner
formula. In a selective measurement, the quasiprobability reduces to
a conditional probability.

It can be mentioned that the Kirkwood distribution may be observed
directly as a statistical average in a successive measurement where
the measurement interaction for the first observable is weak
\cite{Johansen+Mello}.

The Kirkwood distribution has a well-defined operational meaning
both when the interaction with the meter is very strong (i.e. for
projective measurements) and when the interaction is very weak. This
representation also has a well-defined meaning when the experimenter
does not have the possibility of performing maximal measurements.
This could be due to a limited resolution of the meter etc. This is
a relevant situation e.g. in an infinite-dimensional Hilbert space.
In this sense, the Kirkwood distribution may find applications as a
state representation beyond the standard density matrix
representation.

\section{Acknowledgements}

The author is grateful to Pier A. Mello for many interesting and
useful discussions on the foundations of quantum theory in general
and on Kirkwood distribution in particular.


\begin{thebibliography}{40}
\bibitem[{\citenamefont{Johansen}()}]{Johansen-Foot}
One would obtain the same probability if a measurement of the
projector $\hat{B}_n$ were performed. In this case, it would not be
required that the last measurement were of the projective type. It
might as well be e.g. a weak measurement, or for that sake a
measurement of any interaction strength.
\end{thebibliography}
\end{document}